\documentclass[doublecol]{epl2}
\usepackage{amsmath}
\usepackage{amssymb}

\newcommand{\vvr}{{\mathbf{r}}}

\title{Can we always get the entanglement entropy from the Kadanoff-Baym equations? 
The case of the  $T$-matrix approximation.}
\author{M. Puig von Friesen \inst{} \and C. Verdozzi\inst{}  \and C.-O. Almbladh\inst{} }

\institute{
\inst{} Mathematical Physics and European Theoretical Spectroscopy Facility (ETSF), Lund University, 22100  Lund, Sweden
}
\abstract{
We study the time-dependent transmission of entanglement entropy through an out-of-equilibrium model interacting 
device in a quantum transport set-up. The dynamics is performed via the Kadanoff-Baym equations within many-body 
perturbation theory. The double occupancy $\langle \hat{n}_{R \uparrow} \hat{n}_{R \downarrow} \rangle$, 
needed to determine the entanglement entropy, is obtained from the equations of motion of the 
single-particle Green's function. A remarkable result of our calculations is that 
$\langle \hat{n}_{R \uparrow} \hat{n}_{R \downarrow} \rangle$
can become negative, thus not permitting to evaluate the entanglement entropy.
This is a shortcoming of approximate, and yet conserving, many-body self-energies. Among the tested 
perturbation schemes, the $T$-matrix approximation stands out for two reasons: it
compares well to exact results in the low density regime and it always provides a non-negative 
$\langle \hat{n}_{R \uparrow} \hat{n}_{R \downarrow} \rangle$. For the second part of this statement, 
 we give an analytical proof.
Finally, the transmission of entanglement across the device 
is diminished by interactions but can be amplified by a current flowing through the system.
}
\begin{document}
\maketitle
Originally discussed in relation to basic aspects of quantum mechanics,
entanglement is nowadays seen as a key resource for quantum 
information technology \cite{Nielsen}. 
Solid state systems
are well suited to realize quantum information devices,
for their compatibility with conventional electronics hardware.
This has spurred a great deal of cross-disciplinary studies
on entanglement and many-body physics in the condensed phase \cite{Fazio}.

While {\it ab-initio} approaches  to entanglement are starting to be explored \cite{Sham, ida}, 
most studies so far have been for model systems in the ground state
(see e.g. \cite{PeterS,Gu,Larsson&Johannesson1,Capelle,ClaudioPeterConcurrence}). 
However, non-equilibrium studies
are increasing in number (see e.g.\cite{calabrese,Brandes,Kollath,Danieletal}.  
This is because knowing how entanglement is produced/transported is pivotal for
controlled quantum information manipulations, and because it
can give new insight into many-body systems out of equilibrium \cite{Fazio}.\\
\begin{figure}[h]
\begin{center}
\includegraphics[width=8.cm, clip=true]{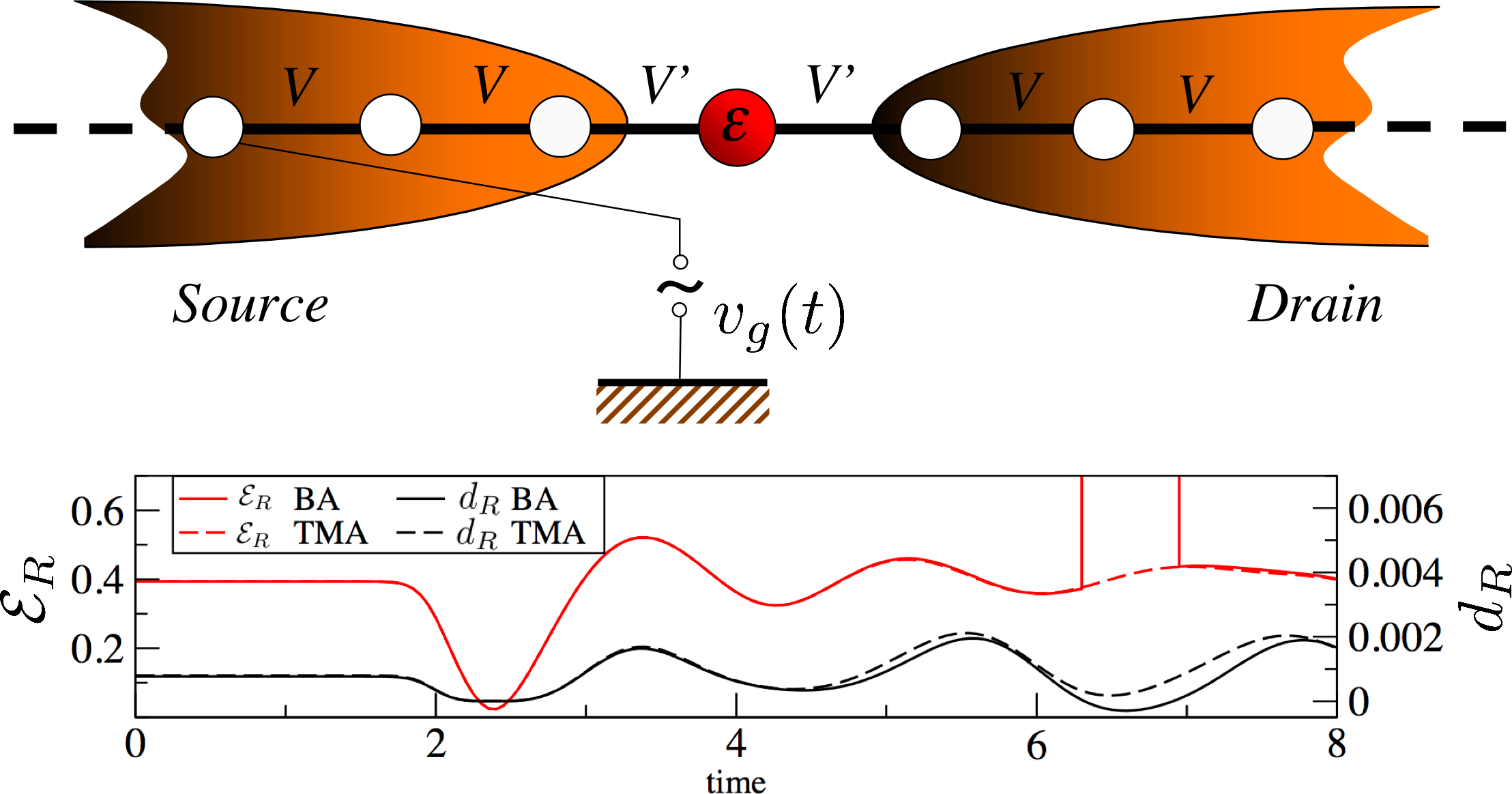}\\
\end{center}
\caption{
(Color online) Top: model dot-leads system. Bottom: Time dependent local entanglement entropy $\mathcal{E}_R$ 
and double occupancy $d_R$, obtained within the $T$-matrix and the second-Born 
approximations (TMA and BA, respectively) for a 7-site isolated cluster (white+red circles) with 2 particles.  
A perturbation $v_g(t)$ has been applied; the system parameters are 
$U=1$, $\epsilon_{0}=-0.5$, $V_0=5$, $T_g=2$, $b_g=0.2$ (see eqs.(\ref{Hlead}-\ref{Hbias})
and main text afterwards). }
\label{docc}
\label{introfigure}
\end{figure}
In this Letter we introduce a non-equilibrium Green's function approach to entanglement,  
using the time-dependent Kadanoff-Baym Equations (KBE) \cite{KadBay,Keldysh}, and
computing a specific measure of entanglement, the entanglement entropy (EE) \cite{Zanardi}, 
within many-body perturbation theory (MBPT). The advantage of our approach is that it can 
be implemented {\it ab initio} \cite{AbinitioKBE}, if one uses reliable
many-body approximations (MBA:s). It is quite natural, at this point, to ask: {\it When are MBA:s 
reliable to obtain the EE ?} 
Here, to address this issue, we choose, as a simple test-case model (fig. \ref{introfigure}), 
an interacting impurity in a quantum transport geometry.

To anticipate the results,
we see (fig. \ref{introfigure}), that, in some time intervals, the EE may diverge,  because the local double occupancy 
(used to obtain the EE) may become negative.   
This non-physical behaviour is a drawback that MBA:s can have
when computing the EE \cite{noteneg}. 
The conditions under which some standard MBA:s can or cannot be used to compute the EE is 
the main outcome of this paper. In particular, we provide an analytical proof of why, among all the
tested MBA:s, the $T$-matrix approximation (TMA, see below) is exempt from the problem shown in fig. \ref{introfigure} \cite{HFApositive}. 
Furthermore, our numerical results show that, compared to exact results in isolated clusters, the TMA yields the most accurate EE.
As an application of our new method to compute the time dependent EE, in the final part of this Letter we will use the KBE 
within the TMA to simulate the time dependent production and transfer of EE in the model junction of fig. \ref{introfigure}, 
and to investigate how the EE is affected by 
interactions and by a current flowing through such a device. 
Before discussing these points in detail, we briefly describe  the model and the method.
\section{The model device} We consider a single interacting impurity (at site 0) coupled to two semi-infinite, 1D leads
(fig.\ref{introfigure}). The Hamiltonian of this system is given by
\begin{equation}
H=H_{imp} + \sum_{\alpha=S,D}[H_\alpha+V_{\alpha,imp}+W_\alpha(t)]+V_{g}(t).\label{Htot}
\end{equation}
The impurity Hamiltonian is $H_{imp}=\epsilon_0 \sum_\sigma\hat{n}_{0\sigma}+U \hat{n}_{0\uparrow}\hat{n}_{0\downarrow}$,
where $\epsilon_0$ and $U$ are the impurity on-site and interaction energies, respectively, and
$\hat{n}_{0\sigma}=a_{0\sigma}^{\dagger}a_{0\sigma}$, with $\sigma=\uparrow,\downarrow$. The following three contributions
describe the leads (source, $S$, and drain, $D$), their coupling to the impurity and 
the effect of an external bias $w_\alpha(t)$:
\begin{eqnarray}
\!\!\!\!H_\alpha\!\!\!&=&\!\!\!\! \sum_\sigma \sum_{i_{\alpha}=1}^{\infty}\epsilon_\alpha \hat{n}_{i_\alpha, \sigma}
-V\!\left(a^\dagger_{i_\alpha, \sigma}a_{i_\alpha+1, \sigma}+H.c.\!\right) \label{Hlead}\\ 
\!\!\!\!V_{\alpha,imp}\!\!&=&\!\!\!-V' \sum_{\sigma} a^{\dagger}_{1_{\alpha},\sigma} a_{0,\sigma} +H.c. \label{Hcontact}\\
\!\!\!\!W_\alpha(t)\!\!\!&=&\!\!\! w_\alpha(t) \sum_\sigma \sum_{i_{\alpha}=1}^{\infty}\hat{n}_{i_\alpha, \sigma}. \label{Hbias}
\end{eqnarray}
The terms $w_\alpha(t)$ and $\epsilon_\alpha$ are the  
time dependent bias and time independent on-site energy in the $\alpha$-th lead, 
(both are taken constant in space in each lead), whereas the $V$ and $V'$ are the 
hopping parameters in the leads and in between lead and impurity respectively.  
The time dependent gate voltage is applied on the third site 
in the $S$ lead, $V_{g}(t) = v_{g}(t)\sum_\sigma \hat{n}_{3_S,\sigma}$. 
$U, \epsilon_0, V', w_\alpha(t), \epsilon_\alpha$ and $v_g(t)=V_0\exp\left(-(\frac{t-Tg}{b_g})^2\right)$ are given in units of $V=1$. 
We take spin-up and -down electrons equal in number;
this holds at all times ($H$ has no spin-flip terms).  

We will also compare EE results from MBPT to exact ones in small clusters; for this,
we consider a finite-size version of the system in fig. {\ref{introfigure}, keeping in each lead only the three
sites closest to the impurity. The Hamiltonian of this seven-site cluster will be the same 
as in eqs. (\ref{Htot}, \ref{Hlead}, \ref{Hcontact}, \ref{Hbias}),
but truncated to describe finite, three-site leads. 
\section{Local entanglement entropy} 
For any lattice site $R$ in our system of eq. (\ref{Htot}), we consider the single-site EE and denote it by $\mathcal{E}_R$.
For a given many-body state $|\Psi \rangle$, and in terms of the local von Neumann entropy,  $\mathcal{E}_R$ is given by \cite{Zanardi} 
\begin{equation}
\mathcal{E}_{R}=-\textup{Tr}\left(\rho_{R}\log_{2}\rho_{R}\right),
\label{localentanglementgeneral}\end{equation}
where the reduced time-dependent density matrix, 
$\rho_{R}= \textup{Tr}_{\{R'\}} \rho$, is a contraction of the full density matrix $\rho=|\Psi\rangle \langle \Psi|$,
and $\{R'\}$ denotes the rest of the system, i.e. $R\notin \{R'\}$.

The local EE has been discussed for several systems studied with different methods,
e.g. the Bethe Ansatz  \cite{Gu,Larsson&Johannesson1},
exact diagonalization \cite{Gu,Capelle}, the density-matrix-renormalization-group \cite {Kollath},
path-integral quantum field theory \cite{calabrese}, static \cite{Capelle} and time dependent \cite{Danieletal} 
density functional theory, to mention a few. 

Here we add another entry  to the list, by proposing a KBE approach to determine $\mathcal{E}_R$.
For a non-magnetic system of spin-1/2 fermions, the single-site EE is \cite{Zanardi} 
\begin{eqnarray}
\mathcal{E}_{R}=\!\!\!&-&\!\!\!2\left(\frac{n_{R}}{2}-d_{R}\right)
\log_{2}\left(\frac{n_{R}}{2}-d_{R}\right)\!-\!d_{R}\log_{2} d_{R}\nonumber\\
\!\!\!&-&\!\!\!\left(1\!-\!n_{R}\!+\!d_{R}\right)\log_{2}\left(1\!-\!n_{R}\!+\!d_{R}\right).
\label{localentanglement}\end{eqnarray}
In eq. (\ref{localentanglement}), $n_{R}=\left<\hat{n}_{R\uparrow}+\hat{n}_{R\downarrow}\right>$ is the local single occupancy
(directly obtained from the single-particle $G$ and the KBE),  and $d_{R}=\left<\hat{n}_{R\uparrow}\hat{n}_{R\downarrow}\right>$ 
is the local double occupancy. 
When and how $d_R$ can be obtained 
via the KBE is discussed in the rest of the paper. 
\section{Kadanoff-Baym equations and many-body approximations} 
The KBE determine the time evolution of the 
non-equilibrium, two-time, single-particle Green's function 
$G(1,2)=-i\left\langle T_\gamma\left[\psi_{H}(1)\psi_{H}^\dagger(2)\right]\right\rangle$, where 
1 denote single-particle space/spin and time labels, $r_1 \sigma_1 t_1$ \cite{labels}.
Here, $T_\gamma$ orders the times $t_1,t_2$ on the Keldysh contour 
$ \gamma$ (fig. \ref{method}(left) \cite{Keldysh}), and the field operators are in the Heisenberg picture; the brackets 
$\langle\rangle$ denote averaging over the initial state (or thermal equilibrium).
Showing explicitly only the time labels (matrix notation/multiplication in space 
and spin indexes is adopted), and specializing to time $t_{1}$, we have  
$\left(i\partial_{t_{1}}-h\left(t_{1}\right)\right)G\left(t_{1},t_{2}\right)=
\delta(12)+\int_{\gamma}\Sigma\left(t_{1},t\right)G\left(t,t_{2}\right)dt$. 
Here $h$ is the single-particle Hamiltonian and $\Sigma$, 
the kernel of the integral equation, is the self energy. $\Sigma$ consists generally of two terms: 
i)  $\Sigma_{MBA}$, which accounts for 
the interactions in the device region and is described within a given MBA (see below) and  ii)  $\Sigma_{emb}$, 
describing the effect of having semi-infinite contacts (leads). For non-interacting leads, the
second contribution can be treated exactly with an embedding procedure \cite{datta,antti,RobertHubb}, giving
$\Sigma_{emb} = \sum_\alpha |V'|^2 \tilde{g}_\alpha$, 
where $\tilde{g}$ is the Green's function of the non-interacting, in general biased, unconnected lead $\alpha$ \cite{trick}.  
The initial state is the correlated \cite{Danielewicz} ground state 
(we work at zero temperature, $1/\beta\rightarrow0$), obtained by solving the 
Dyson equation $G=G_{0}+G_{0}\Sigma[G] G$ self consistently, with $(\epsilon-h-\Sigma_{emb})G_0=1$.
In practice we solve these equations using an algorithm discussed in \cite{pva1,pva2}. 

In general, an exact determination of the many-body part $\Sigma$ is not possible,
and one resorts to approximations to include the effects of the interactions in the device 
(hence the name $\Sigma_{MBA}$ given to this part of $\Sigma$).
\begin{figure}[t]
\begin{center}
\includegraphics[width=7.5cm, clip=true]{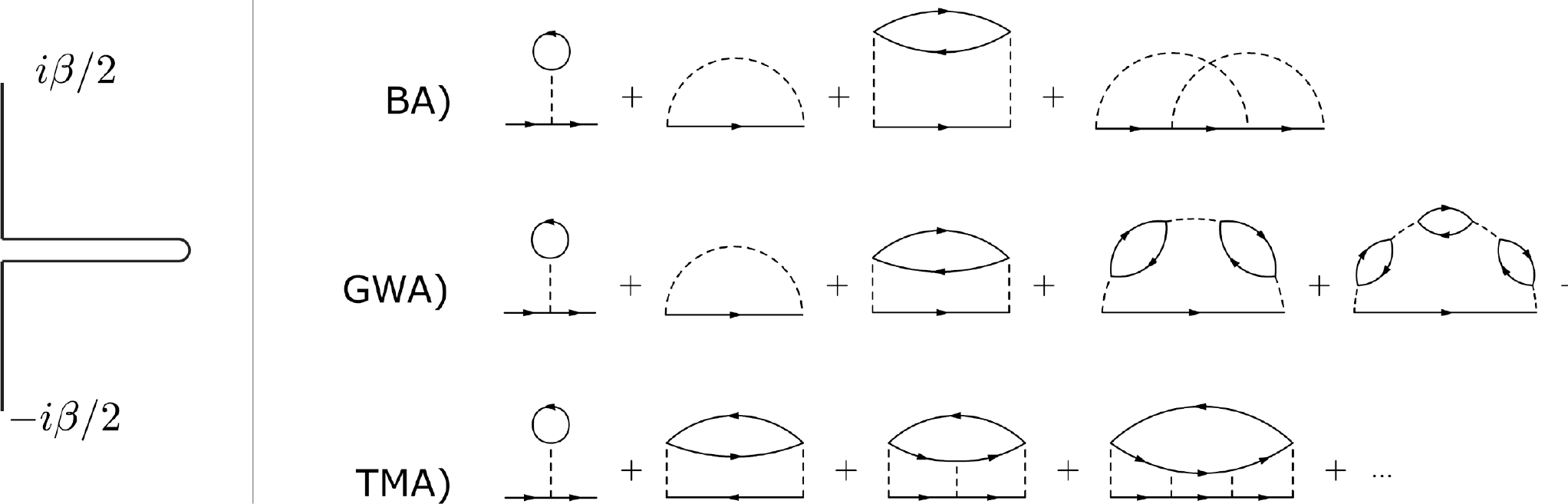}\\
\end{center}
\caption{Contour (left) and approximate self-energies (right).}
\label{method}
\end{figure}
%
Besides the TMA, the other two MBA:s we will consider are the second Born and the $GW$ 
approximations (BA and GWA, respectively) 
(fig. \ref{method}(right)); they are all conserving \cite{KadBay}, i.e. quantities such as the 
total energy and momentum as well as the number of particles are constant during the time evolution. 
The self energy in the BA includes all terms up to second order. 
The GWA \cite{Hedin} amounts to adding up all the bubble diagrams which give rise to the screened 
interaction, $W=U+UPW$, where $P(12)=G(12)G(21)$. In this case the self energy is $\Sigma(12)=G(12)W(12)$.
In a spin-dependent treatment of the TMA \cite{TMA,pva2} 
one constructs the $T$ by adding up all the ladder 
diagrams, $T=\Phi-\Phi U T$, where $\Phi(12)=G(12)G(12)$. 
The expression of the self energy then becomes 
$\Sigma(12)=\int U(13)G(43)T(34)U(42)d34$.
For a detailed account
of the different MBA:s see e.g. \cite{pva2}.
%
\section{Double occupancy from the Kadanoff-Baym Equations:
approximate vs exact results} 
While in principle $G(12)$ gives access only to expectation values
of one-body operators, its equations of motion (i.e., the KBE) permit to obtain also some quantities
which involve expectation values of two-body operators, such as the total energy.
This route can be also used for the double occupancy $d_R$ \cite{Abrikosov};
for on-site ($U_R\neq0$) interactions \cite{interaction},
\begin{equation}
d_R=\left<\hat{n}_{R\uparrow}\hat{n}_{R\downarrow}\right>=
-\frac{i}{U_R}\left[\int_\gamma \Sigma_{MBA}\left(13\right)G\left(31^{+}\right)d3\right]_{RR},
\label{doccformula}\end{equation}
valid for both isolated and contacted (to leads) systems \cite{dimercondmat,gooding}. 
Equation (\ref{doccformula}), together with eq. (\ref{localentanglement})
and the KBE, is the proposed new method to compute the 
dynamical EE, in both model systems calculations and {\it ab-initio}
treatments of real materials \cite{interaction}.

As a benchmark to the method, we considered $\mathcal{E}_R$ for an isolated cluster (the seven-site system explicitly 
shown in fig. \ref{introfigure}) and computed  eq. (\ref{localentanglement}) i) exactly and
ii) using eq. (\ref{doccformula}) and the KBE+MBA:s scheme.
\begin{figure}[t]
\begin{center}
\includegraphics[width=7.7cm, clip=true]{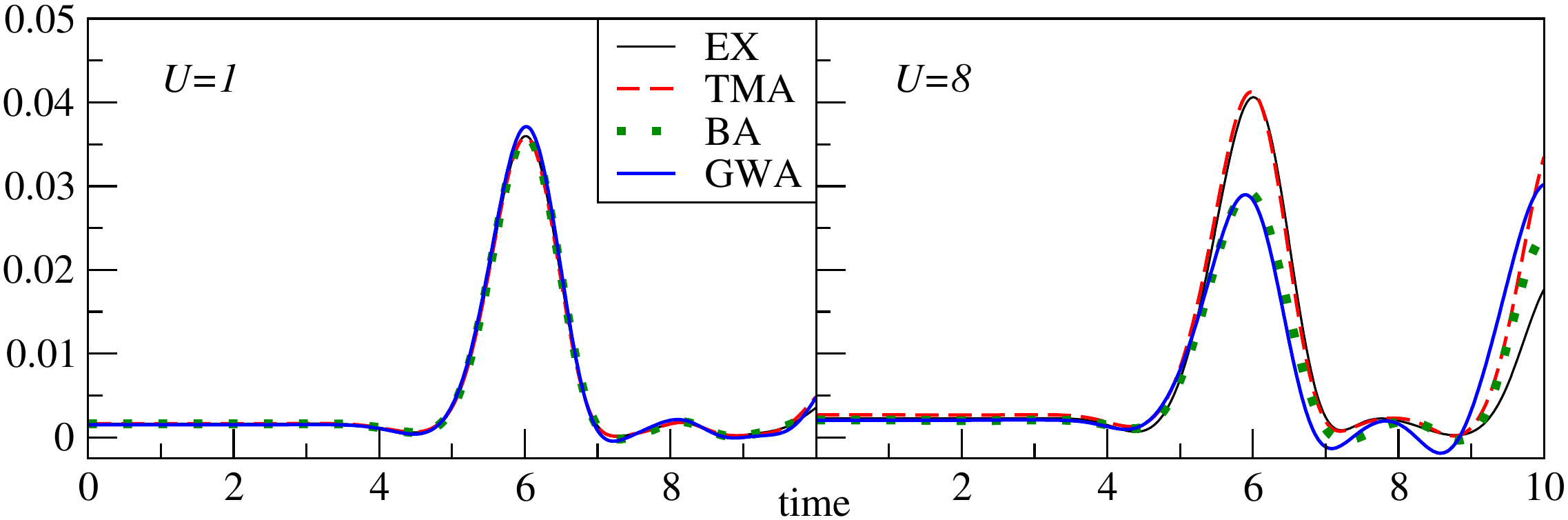}\\
\end{center}
\caption{(Color online) Time-dependent double occupancy on the 7th site for the 7-site 
cluster with $n=1/7$, $T_g=2$, $V_0=5$, $b_g=0.2$. 
The curves correspond to exact (thick black solid), TMA (dashed red), BA 
(dotted green) and GWA (thin solid blue). In (a) $U=1$ and in (b) $U=8$.}
\label{doubleocc}
\end{figure}
The numerical results for $d_R$ are presented in fig. \ref{doubleocc},
for the exact, BA, GWA and TMA treatments, in the low
density regime. Two values of the interaction are considered, and $d_R$ 
is calculated for the last (i.e. the seventh) site of the cluster. As anticipated in fig. \ref{introfigure},
and also shown
for a Hubbard dimer in \cite{noteneg}, the TMA is the only scheme,
among those considered, to produce correlation functions always positive.
Moreover we notice the quite good agreement between the TMA and
the exact $d_R$:s at low densities (higher densities, near half-filling,  
will be discussed later in the paper). These results clearly indicate that one
should abandon the use of the BA and the GWA to compute
the EE: in what follows, we will only consider the TMA.
However, before analyzing the EE in  the TMA, we 
describe in some detail why the TMA local double 
occupancies are always non-negative.
%
\section{Positiveness of $T$-matrix} 
All the MBA:s used in this work are conserving. This, however,
does not guarantee that other properties (e.g. spectral 
features \cite{COAspectra, Holm} or response functions) obtained with
such MBA:s automatically
satisfy further basic criteria. In fact, in fig. \ref{docc}, 3 we saw that 
density and pair correlations for GWA and BA 
may violate the positivity condition
\begin{eqnarray}
\langle \psi^\dagger(1) \psi^\dagger(2) \psi(2) \psi(1) \rangle \ge 0 .
\label{eq:positive}
\end{eqnarray}
The pair correlation function
is actually a rather sensitive measure of the quality of a many-body 
approximation. It has been known for a long time that the 
random-phase approximation (RPA) \cite{Lindhard}
in the ground state
gives pair correlations which, at metallic densities,  are strongly negative 
at short distances \cite{Alf}.  Subsequently, negative pair correlations at short 
distances were also found  within the self-consistent GWA \cite{Holm}.

One may ask if the situation would improve if the response function was
calculated self-consistently from \textit{changes} in $\Sigma$ and 
thereby via the Bethe-Salpeter equation:  Such a response function fulfills 
all macroscopic conservation laws, but does not necessarily yield pair 
correlation functions everywhere positive. For example, the RPA discussed above
is the response function in the Kadanoff-Baym sense of the conserving
Hartree approximation but it violates the positiveness condition in eq. (\ref{eq:positive}) \cite{Alf}.

According to figs. \ref{introfigure}(b,c) and \ref{entagcluster}, the TMA, among the examined MBA:s, 
is the only one giving always positive correlation functions
(and, in many cases in good agreement  with exact cluster results).
This is not due to fortuitously chosen parameter values: 
we will now give proof that the pair correlation from the TMA 
is manifestly positive, starting with the equilibrium case.

In the ground state, for the TMA, the pair correlation function
is a simple contraction of the $T$-matrix itself,
\begin{eqnarray}
\langle \hat{n}_{R\uparrow} \hat{n}_{R\downarrow}\rangle = -i T_{RR}(t, t^+)
= -i T^<_{RR}(t, t),
\label{eq:T_mat_corr_hubb}
\end{eqnarray}
and its positiveness is a consequence of the positiveness
of the $T$-matrix spectral function. 
Here $>$ ($<$) refers to the electron (hole) part. 
The basic 
building block in the TMA is $\Phi_{RR^\prime}(t-t^\prime)  = -i G_{RR^\prime}(t-t^\prime) 
G_{RR^\prime}(t-t^\prime)$. In Fourier space we have
\begin{eqnarray}
\Phi_{RR^\prime}(\epsilon) = \int \frac{C_{RR^\prime}(\epsilon^\prime) d\epsilon^\prime}
{\epsilon - \epsilon^\prime + i \eta\; sgn(\epsilon^\prime - 2 \mu)} ,
\label{eq:C_spec}
\end{eqnarray}
where the spectral function $C_{RR^\prime}$ is given by 
\begin{eqnarray}
C_{RR^\prime}(\epsilon) =  \int_{\epsilon - \mu}^\mu 
A_{RR^\prime}(\epsilon^\prime) A_{RR^\prime}(\epsilon - \epsilon^\prime) d\epsilon^\prime. 
\nonumber
\end{eqnarray}
\noindent Thus, $C_{RR^\prime}(\epsilon)$ is positive (negative) definite for $\epsilon < 2 \mu$ 
($\epsilon > 2 \mu$).
The Dyson equation for the $T$-matrix is
$ \hat{T}(\epsilon) = \hat{\Phi}(\epsilon) - \hat{\Phi}(\epsilon)\hat{U}
\hat{T}(\epsilon) $
(here $\hat{\Phi}, \hat{T}, \hat{U}$ are matrices in one-particle indices \cite{matrix}, 
and $U_{RR'}=U_R\delta_{RR'}$). This gives
\begin{eqnarray*}
&&\hat{T}^\lessgtr = \hat{\Phi}^\lessgtr - \hat{\Phi}^\lessgtr \hat{U} \hat{T}^a -
\hat{\Phi}^r \hat{U} \hat{T}^\lessgtr , \\
&&(1 + \hat{\Phi}^r \hat{U} ) \hat{T}^\lessgtr = \hat{\Phi}^\lessgtr (
1 - \hat{U} \hat{T}^a ),
\end{eqnarray*}
and
\begin{eqnarray}
\hat{T}(\epsilon) - \hat{T}^\dagger(\epsilon)
= [1 - \hat{T^r}(\epsilon) \hat{U} ] 
[ \hat{\Phi}(\epsilon) - \hat{\Phi}^\dagger(\epsilon) ]
[1 - \hat{U} \hat{T}^a(\epsilon) ],
\nonumber
\end{eqnarray}
where we have used the identity
$(1 - \hat{U} \hat{T} )  (1 + \hat{\Phi} \hat{U}) = 1$,
and $a,r$ have their usual meaning.
The $T$-matrix has thus the spectral decomposition
\begin{eqnarray}
\hat{T}(\epsilon) = \int \frac{\hat{D}(\epsilon^\prime) d\epsilon^\prime}
{\epsilon - \epsilon^\prime + i \eta\; sgn(\epsilon^\prime - 2 \mu)}
\label{eq:T_D}, 
\end{eqnarray}
where  $\hat{D}(\epsilon) = [1 - \hat{T}^r(\epsilon) \hat{U} ] \hat{C}(\epsilon)
[1 - \hat{U} \hat{T}^a(\epsilon) ]$. 
Consequently, $\hat{D}(\epsilon)$ is positive definite for $\epsilon < 2 \mu$,
and
\begin{eqnarray}
\langle \hat{n}_{R\uparrow} \hat{n}_{R\downarrow}\rangle = -i T_{RR}(t, t^+)
= \int_{-\infty}^{2 \mu} D_{RR}(\epsilon^\prime)d\epsilon^\prime  \ge 0 .
\label{eq:pos_T_hubb}
\end{eqnarray}
\noindent The above result remains valid for a general two-body interaction
$u(\vvr_1 - \vvr_2) \delta(t_1 - t_2)$, provided we use 
a symmetrised TMA which includes both direct and exchange ladder
diagrams. In this case, the basic entities are
\begin{eqnarray*}
&&\Phi(\vvr_1 \sigma_1, t_1, \vvr_2, \sigma_2, t_1; \vvr_3 \sigma_3, t_3, \vvr_4, \sigma_4,t_3) 
\nonumber \\
&&= \langle \vvr_1 \sigma_1, \vvr_2, \sigma_2 | \hat{\Phi}(t_1, t_3) | \vvr_3 \sigma_3, \vvr_4, \sigma_4 \rangle 
\nonumber \\
&&= -i \left \{ G(13) G(24) - G(14) G(23) \right \}_{t_2= t_1=t, t_4 = t_3=t^\prime},
\end{eqnarray*}
and $\hat{T}(t, t^\prime)$, which can be considered as matrices in a complete set
of two-particle labels and depending on two times $t, t^\prime$\cite{labels}.
Specializing to the ground state, and in Fourier space,
\begin{eqnarray}
\hat{T}(\epsilon) = \hat{\Phi}(\epsilon) -\frac{1}{4}
\hat{\Phi}(\epsilon) \hat{\Gamma}^{(0)} \hat{T}(\epsilon) .
\label{eq:T_Dyson}
\end{eqnarray}
Here,$
\Gamma^{(0)}(12;34) = u(12) [ \delta(13)\delta(24) - \delta(14) \delta(23) ]
$
is a symmetrized interaction. Finally, for the self energy,
\begin{align}
(\Sigma G)(\vvr_1, t_1, \sigma_1; &\vvr_2, t_2, \sigma_2) =
\!-i\!\!\! \int\!\!\! u(13)\langle T_\gamma \hat{n}(3) \psi(1) \psi^\dagger(2) \rangle d3
\nonumber \\
&= - \int u(13) T(1 3^+  ; 2 3^{++}) d3.\label{EOMTM}
\end{align}
In eq. (\ref{EOMTM}), the first equality follows from the equation 
of motion for $G$, and the second defines the symmetrised TMA.
The correlation part of $\Sigma$ is shown diagrammatically in fig. \ref{fig:TMAs}.
\begin{figure}[t]
\begin{center}
\includegraphics[width=7.7cm, clip=true]{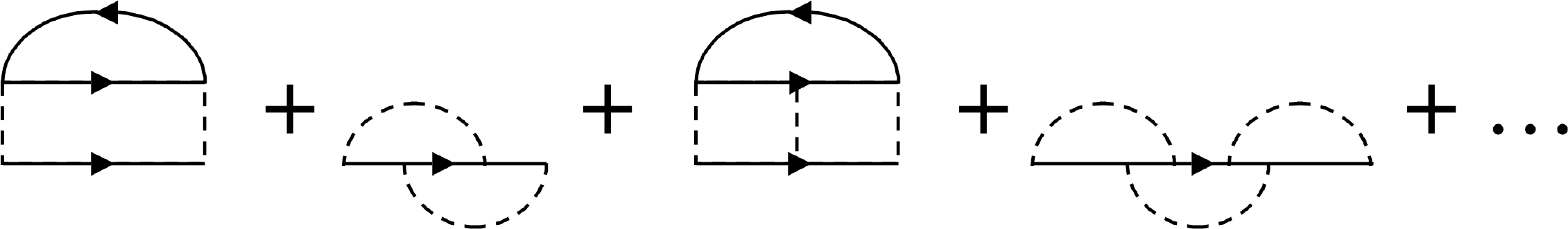}\\
\end{center}
\caption{Correlation part of the symmetrized TMA self energy.}
\label{fig:TMAs}
\end{figure}
Assuming the pair interaction invertible, the equal-time 
pair correlation is given by the spin contraction of
\begin{eqnarray}
\langle \psi^\dagger(1) \psi^\dagger(2) \psi(2) \psi(1) \rangle =
- i \langle 12 | \hat{T}(t, t^+) | 12 \rangle .
\label{eq:T_mat_corr2}
\end{eqnarray}
The very construction of the TMA in terms of multiple pair scatterings suggests
that the individual spin components
in eq. (\ref{eq:T_mat_corr2}) can be interpreted as approximations to spin-decomposed
correlations. Thus, the close correspondence between the contracted $T$-matrix
and the pair correlations remains also with a general interaction,
cf. eqs. (\ref{eq:T_mat_corr2}, \ref{eq:T_mat_corr_hubb}). In Fourier space, the matrix 
$\hat{\Phi}$ has a spectral decomposition as in eq. (\ref{eq:C_spec}), where 
in this case
\begin{equation}
\langle 12 | \hat{C}(\epsilon)| 34 \rangle \!=\!\! \int_{\epsilon - \mu}^{\mu}\!\!\!\!\! 
[ A_{13}(\epsilon') A_{24}(\epsilon - \epsilon') 
- A_{14}(\epsilon') A_{23}(\epsilon - \epsilon') ]
d\epsilon',
\nonumber
\end{equation}
and is positive definite for $\epsilon < 2 \mu $. From the Dyson 
equation for $\hat{T}$ (cf. eq. \ref{eq:T_Dyson})
we then obtain
\begin{eqnarray}
\hat{T}(\epsilon) \!-\! \hat{T}^\dagger(\epsilon) \!=\! [1 \!-\! \frac{1}{4}\hat{T}(\epsilon) \hat{\Gamma}^{(0)}]
[\hat{\Phi}(\epsilon) \!-\! \hat{\Phi}^\dagger(\epsilon)]
[1 \!-\! \frac{1}{4}\hat{\Gamma}^{(0)} \hat{T}^\dagger(\epsilon)].\nonumber
\label {eq:lang1}
\end{eqnarray}
Thus $\hat{T}$ has a spectral decomposition similar to that of 
$\hat{\Phi}$, and its spectral function $\hat{D}$ is positive definite
for $\epsilon < 2 \mu$.
The correlation function can now be expressed in terms of the diagonal 
element of $\hat{D}$,
\begin{eqnarray}
\langle \psi^\dagger(1) \psi^\dagger(2) \psi(2) \psi(1) \rangle =
- i \langle 12 | \hat{T}(t, t^+) | 12 \rangle 
\nonumber \\
= \int_{-\infty}^{2 \mu} \langle 12 | \hat{D}(\epsilon^\prime) | 12 \rangle d\epsilon^\prime \ge 0 .
\;\;\;\;\;\;\;\;
\label{eq:pos_T}
\end{eqnarray}
Thus, the spin-decomposed
pair correlations are positive and thereby also their spin averages.

Such result remains valid in the presence of an external field for $t > 0$.
The proof is similar to the proof that manifestly positive 
$\mp i \Sigma^\lessgtr$ give manifestly positive $\mp i G^\lessgtr$ via
the KBE.
To simplify the analysis we stay at zero temperature and deform the contour in 
fig. \ref{method}(left) into one which runs from $- \infty$ to $\infty$ and back 
again (the equivalence follows from the
analyticity properties of greater/lesser functions when Re$\;t \leq 0$ and thus 
before any external field is applied). With this contour, the
$T$-matrix equation becomes
\begin{eqnarray}
\tilde{T} = \tilde{\Phi} -\frac{1}{4}
\tilde{\Phi} \tilde{\Gamma}^{(0)} \tilde{T} ,
\label{eq:T_Dyson2}
\end{eqnarray}
where now the matrix labels include also time,
\begin{eqnarray*}
\langle t, \vvr_1, \sigma_1, \vvr_2, \sigma_2 | \tilde{X} 
| t^\prime, \vvr^\prime_1, \sigma^\prime_1, \vvr^\prime_2, \sigma^\prime_2 \rangle = \\
X(t_1, \vvr_1 \sigma_1, \vvr_2, \sigma_2; t_1^\prime, \vvr^\prime_1, \sigma^\prime_1,
\vvr^\prime_2, \sigma^\prime_2) .
\end{eqnarray*}
This leads to
$\tilde{T}^< = 
[1 - \frac{1}{4}\tilde{T}^r \tilde{\Gamma}^{(0)}]
\tilde{\Phi}^< [1 -\frac{1}{4} 
\tilde{\Gamma}^{(0)} \tilde{T}^a ]$.
Any positive-definite $G^<$ gives a positive-definite $\tilde{\Phi}$,
and because $\tilde{T}^r$ is the Hermitian conjugate of $\tilde{T}^a$,
$\tilde{T}^<$ is manifestly positive, and produces
positive pair correlations \cite{finiteT}. As a side remark, we could add 
that the BA, GWA, and TMA all give
manifestly positive spectral densities in equilibrium and positive
$\pm G^{\lessgtr}$ out of equilibrium, a condition which not all conserving
schemes obey.
%
\section{Entanglement propagation across a model device} 
\begin{figure}[t]
\begin{center}
\includegraphics[width=7.7cm, clip=true]{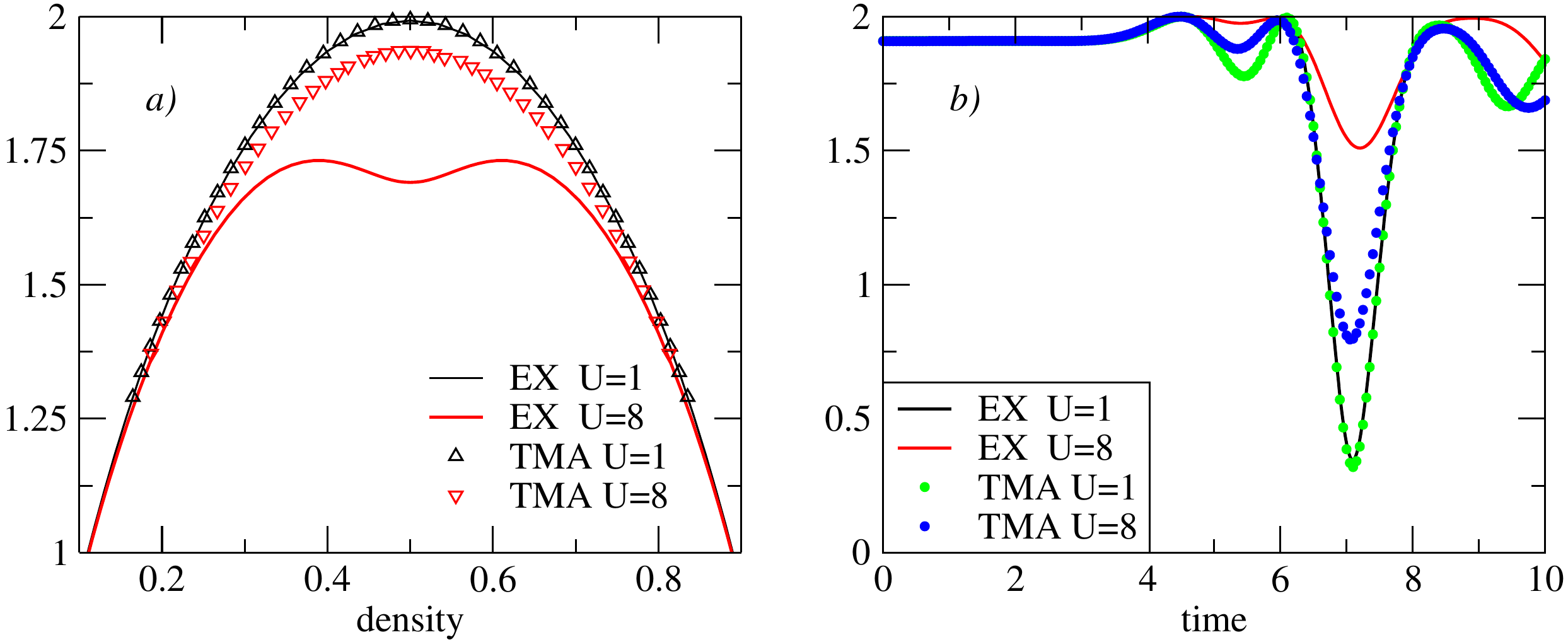}\\
\end{center}
\caption{(Color online) Groundstate (a) and time dependent (b) entanglement for the 7-site 
cluster on the 4th (a) and 7th (b) sites, with $n=1/7$ for $U=1,8$. 
In (b), $T_g=2$, $V_0=5$, $b_g=0.2$ and $\epsilon_0=-U/2$. The curves 
correspond to exact (solid) and TMA (triangles, dots).}
\label{entagcluster}
\end{figure}

Since double occupancies cannot be negative in the TMA,
we now use this approximation  to 
investigate the production, offline-transmission and monitoring of EE, following the onset of an external
gate voltage $V_g(t)$ (fig. \ref{introfigure}).
In the regimes considered, correlation damping and metastable states play an insignificant role \cite{pva1,pva2}.

In the low density regime, the TMA works well also quantitatively \cite{pva1}. Thus, a suitable regime for our simulations
is when the density impurity level in the cluster geometry of fig. \ref{introfigure}, both in the ground state and during the dynamics,
has low occupations. In that regime, the difference between EE results obtained with 
small and large $U$ values was found to be negligible. 

This outcome can be understood looking at fig. \ref{entagcluster}a, where we 
show, as function of the impurity density, the ground state EE for different $U$:s at the central site of our 7-site cluster. To produce the curves in fig. \ref{entagcluster}a, the density is varied continuously by tuning the on-site energy. We see that the dependence of the EE on the interaction strength is only significant in the half-filling regime \cite{Capelle1}.  This remark readily applies to the time-dependent case, since we are computing the EE dynamics 
within an adiabatic local density approximation \cite{Danieletal} which uses eq. (\ref{localentanglement}).

We  then consider an impurity density near half-filling.
The inherent time dependent results are shown in fig. \ref{entagcluster}b, where it is immediately apparent that, as expected from the entanglement diagram of fig. \ref{entagcluster}a, electron correlations considerably affect transmission. One also notes
that while becoming quantitatively inaccurate, the TMA reproduces in a qualitatively correct 
way the exact EE.
\section{The effect of an electric current}
According to the EE diagram in fig. \ref{entagcluster}a, if the occupation at the 4th site in the cluster is close to half filling, the EE is close to being maximal.
Hence, changes in density at that site (except for variations limited to the small region around $n=1$, where the EE may have a local minimum) can only reduce the EE. The other sites in the cluster exhibit the same qualitative behavior.
Thus, when the isolated 7-cluster device is close to half filling in the ground state (as in fig. \ref{entagcluster}b),
a voltage pulse on the left of the device will produce a local reduction of the EE. At a later time, this will induce a change
of density to the right of the device, which also causes a local decrease of EE. This is the behavior observed in fig. \ref{entagcluster}b,
i.e. the left-to-right transmission of a local {\it depression} in the EE, which is sensitive to the strength of the interaction $U$ at the impurity site. 
\begin{figure}[t]
\begin{center}
\includegraphics[width=7.7cm, clip=true]{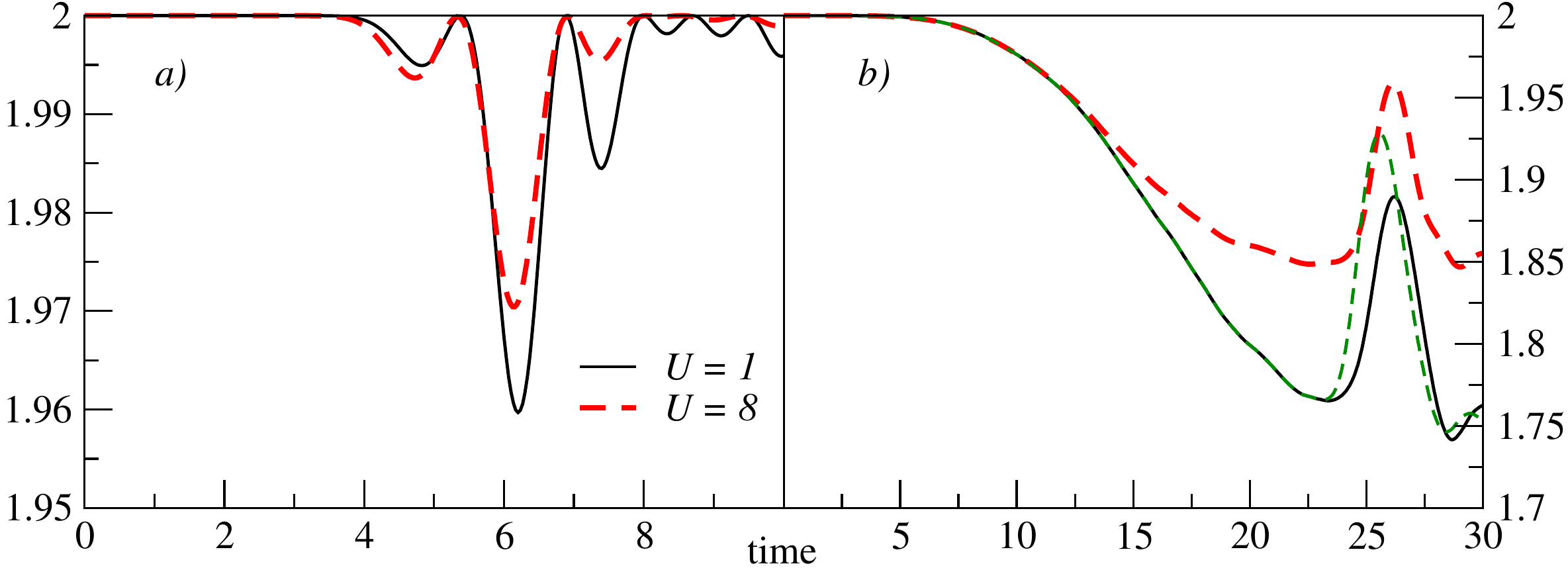}\\
\end{center}
\caption{ (Color online) Time dependent entanglement on 7th site. $V_0=5$, $b_g=0.2$, $U=1,8$ and $\epsilon_0=-U/2$.
In (a) the leads are unbiased with $T_g=2$ and in (b) they are biased, 
$w_{S/D}=\pm V_b{\rm sin}(\frac{\pi}{2T_b}t)$, where $V_b=1$, $T_b=20$ and $T_g=22$. The green dashed
line corresponds to $V_b=-1$.
}
\label{entagtrans}
\end{figure}

For a contacted device, one has an additional handle, the electric current, to choose the most suitable 
initial point (interval) in the EE diagram, and thus to tune the "EE background" on which 
a voltage pulse can be superimposed. This is illustrated in fig. \ref{entagtrans}.
In the left panel (a), we display the results for an unbiased system (no current) at half-filling.
Here, after the pulse is applied, the mechanism for EE propagation is the same as in fig.
 \ref{entagcluster}b, and we observe the transmission of a local depression in the EE.
To provide a simple illustration of  the effect a current may have,  we applied a bias and evolved the system (fig. \ref{entagtrans}b)
semi-adiabatically to approach the steady-state current and EE, at $t\approx 20$, before applying a pulse $V_g(t)$.
In this case, the transmitted EE pulse {\it increases} in amplitude from its steady state value. For $U=1$, if the bias is reversed the pulse arrives faster, 
increases in size and gets broadened. This effect is, though not expected, of simple
single-particle nature.
\section{Conclusions}
We presented a new method to obtain the time-dependent local entanglement entropy, via the Kadanoff-Baym
equations and within MBPT. The method can be implemented {\it ab-initio}, i.e. for realistic materials.
We showed that for some commonly used MBA:s (specifically, the BA and the GWA), the resulting double occupancy 
may become negative, i.e. leading to 
a diverging entanglement entropy. We provided a proof that this is never the case for the TMA:
the latter always gives positive double occupancy and, for finite systems, good agreement with 
exact numerical results in the low density regime.
Finally, within the KBE+TMA scheme, using
a model interacting device in a quantum transport setup, we illustrated the role played
by the current in altering the modality of entanglement transmission across the device.
\acknowledgments
We acknowledge Ulf von Barth and Daniel Karlsson for critical reading the manuscript. This work was supported by the 
EU 6th framework Network of Excellence NANOQUANTA (NMP4-CT-2004-500198) and the 
ETSF facility (INFRA-2007-211956).


\begin{thebibliography}{10}
\bibitem{comfig1} 
The acronyms in fig. \ref{introfigure} as well as the parameters in the caption are defined below.
\bibitem{HFApositive} The Hartree-Fock approximation yields positive definite
double occupancies as a trivial consequence for not incorporating 
any correlation effects and will thus not be considered in this letter.
\bibitem{labels}
The labels 1 and 12 denote the one and two-particle space/spin labels, respectively: $1\equiv\vvr_1 \sigma_1$ and
$12\equiv (\vvr_1 \sigma_1 \vvr_2 \sigma_2)$.
 In a few places, 1 will also include a time argument: $1 \equiv (\vvr_1,\sigma_1,t_1)$; this will be clear from the context.
In lattice models, we denote $\vvr$ by $R$.
\bibitem{trick} 
The embedding scheme is valid only for a space independent (in general time varying)
on-site energy. However, the gate voltage, $V_g(t)$, is applied on the third 
site of the $S$ lead. To reconcile this with the embedding technique we enlarge 
the central region to include all sites with space 
dependent fields.
\bibitem{dimercondmat} A preliminary discussion of eq. (\ref{doccformula})  can be found in 
\cite{noteneg}.
\bibitem{matrix} 
In general the $\hat{T}$ and $\hat{\Phi}$ can be seen as matrices in two-particle space but
for on-site interaction they reduce to matrices in single-particle labels.
\bibitem{Capelle1} The qualitative shape of the EE in the left panel of fig. \ref{entagcluster}
has been also observed in other systems and with other methods, e.g. using lattice
Density Functional Theory  for the homogeneous 1D Hubbard model \cite{Capelle}.
\end{thebibliography}
\end{document}